\documentstyle[graphicx,amsfonts,amssymb,aps,prd]{revtex}

\draft

\begin{document}

\title{Hydrodynamic approach to the evolution of cosmological
  structures}

\author{Alvaro Dom\'\i nguez\footnote{email:
    alvaro@theorie.physik.uni-muenchen.de}}

\address{Theoretische Physik, Ludwig--Maximilians--Universit\"at,
  Theresienstr. 37, D--80333 M\"unchen, Germany}

\address{Laboratorio de Astrof\'\i sica Espacial y F\'\i sica
  Fundamental, Apartado 50727, E--28080 Madrid, Spain}


\maketitle

\vspace{\baselineskip}
\centerline{Phys. Rev. D; submitted: May 18, 2000; accepted: July 18, 2000}

\begin{abstract}
  A hydrodynamic formulation of the evolution of large-scale structure
  in the Universe is presented. It relies on the spatially
  coarse-grained description of the dynamical evolution of a many-body
  gravitating system. Because of the assumed irrelevance of
  short-range (``collisional'') interactions, the way to tackle the
  hydrodynamic equations is essentially different from the usual case.
  The main assumption is that the influence of the small scales over
  the large-scale evolution is weak: this idea is implemented in the
  form of a {\it large-scale expansion} for the coarse-grained
  equations. This expansion builds a framework in which to {\em
    derive} in a controlled manner the popular ``dust'' model (as the
  lowest-order term) and the ``adhesion'' model (as the first-order
  correction).  It provides a clear physical interpretation of the
  assumptions involved in these models and also the possibility to
  improve over them.
\end{abstract}

\pacs{98.80.-k, 98.80.Hw, 98.65.Dx, 04.40.-b}

\section{Introduction}

The standard model to understand the large-scale features of the
matter distribution in the Universe after decoupling from radiation
can be hardly simpler: a collection of many identical point particles
interacting with each other via the Newtonian gravitational force in
an expanding spatial background \cite{Peeb80,Padm93}. Structure arises
as a consequence of the gravitational instability of initially tiny
density perturbations. This model neglects relativistic effects, which
become important only at scales of the order of the horizon and
beyond, or when dealing with relativistic velocities. The model also
excludes nongravitational interactions, which are assumed to be
relevant only at small enough scales.

The general solution to the dynamical evolution of this model is
unknown due to the mathematical difficulties. N-body simulations,
which numerically solve the dynamical equations (see
Eqs.\ (\ref{newton}) in the next section), have been very helpful in
understanding this evolution. In this work I look for an analytical
derivation of some of the relevant features in the formation of
large-scale structures. One is not usually interested in following
the detailed path of each particle, but rather in some general
properties that typically depend on the behavior of a large number of
particles and which in the end are the kind of data provided by
observations, e.g., the smoothed density and velocity fields. This
naturally leads to the use of a {\em coarse-grained} or smoothed
description: the evolution of a few, ``macroscopic'' variables is
isolated by making suitable approximations for the evolution of the
neglected degrees of freedom and for their influence on the relevant
ones.

The idea of a spatial coarse-graining has a long history in the field
of statistical mechanics, where it has proven quite successful as a
powerful tool for extracting information about the dynamical evolution
of many-body systems. In this work I explore its application in the
context of cosmological structure formation. Although the procedure of
smoothing a field is widely used in cosmology, this application has
mainly had a descriptive purpose.  Unlike this, I study systematically
the {\em dynamical} evolution of the smoothed fields.  As we shall
see, this method has the merit of providing a common framework for
different models (dust and adhesion) of structure formation: it offers
a clear explanation of the approximations involved in each model and
of their physical meaning, so that it also opens the way to
systematically relax them and obtain improved models.  Starting from
the microscopic equations of motion of the particles, I derive an
exact set of hydrodynamic-like equations for the (coarse-grained)
mass-density and velocity fields (Sec.\ \ref{seccoarse}). These
equations do not form an autonomous set, but rather constitute the
first ones of an infinite hierarchy that must be truncated to become a
useful tool. I discuss in Sec.\ \ref{secsmallk} how this can be
achieved by resort to a {\em large-scale expansion}, based on the
assumption that the dynamical influence of the small scales over the
large-scale evolution is weak.  The lowest-order term of the expansion
yields the dust model (Sec.\ \ref{secdust}), and it corresponds to a
complete neglect of the structure below the coarsening scale. This
leads to an eventual failure of the model (in the form of pancake-like
singularities), which can be prevented by considering the first-order
correction in the expansion (Sec.\ \ref{secadhesion}): it accounts for
the dynamical influence of the structure below the coarsening scale,
and I show by means of boundary-layer techniques that it gives rise to
a robust ``adhesive'' behavior of the same kind as the adhesion model.
I end up in Sec.\ \ref{secconclusion} with a discussion of the
results.

\section{Coarse-graining the basic equations} 
\label{seccoarse}

The basic model is a system of nonrelativistic, identical point
particles which (i) are assumed to interact with each other via
gravity only; (ii) look homogeneously distributed on sufficiently
large scales, so that the evolution corresponds to an expanding
Friedmann-Lema\^\i tre cosmological background; and (iii) deviations
to homogeneity are relevant only on scales small enough that a
Newtonian approximation is valid to follow their evolution. Let $a(t)$
denote the expansion factor of the Friedmann-Lema\^\i tre cosmological
background, $H(t) = \dot{a}/a$ the associated Hubble function, and
$\varrho_b (t)$ the homogeneous (background) density on large scales.
${\bf x}_i$ is the comoving spatial coordinate of the $i$-th particle,
${\bf u}_i$ is its peculiar velocity, and $m$ its mass. In terms of
these variables the evolution is described by the following set of
equations \cite{Peeb80} ($\nabla_i$ denotes a partial derivative with
respect to ${\bf x}_i$):
\begin{mathletters}
  \label{newton}
  \begin{equation}
    \dot{\bf x}_i = {1 \over a} {\bf u}_i , 
  \end{equation}
  \begin{equation}
    \dot{\bf u}_i = {\bf w}_i - H {\bf u}_i ,
  \end{equation}
  \begin{equation}
    \nabla_i \cdot {\bf w}_i = - 4 \pi G a \left[ {m \over a^3} 
      \sum_{j \neq i} \delta({\bf x}_i - {\bf x}_j) - \varrho_b \right] , 
  \end{equation}
  \begin{equation}
    \nabla_i \times {\bf w}_i = {\bf 0} ,
  \end{equation}
\end{mathletters}
where ${\bf w}_i$ is the peculiar gravitational acceleration acting on
the $i$-th particle. Finally, Eqs. (\ref{newton}) must be subjected
to periodic boundary conditions in order to yield a Newtonian
description consistent with the Friedmann-Lema\^\i tre solution at
large scales \cite{BuEh97}.

To implement the idea of a spatial coarse-graining one employs a
smoothing window $W(z)$: this smoothing window should define a bounded
region of space and whose inner structure is smoothed out. In App.\ 
\ref{apwindow} I discuss the general properties I will require from a
smoothing window.  Let $L$ denote the (comoving) coarse-graining
length scale. A microscopic mass density field and a coarse-grained
mass density field can be defined respectively as follows:
\begin{mathletters}
  \label{density}
  \begin{equation}
    \varrho_{mic} ({\bf x}, t) = {m \over a(t)^3} \sum_i 
    \; \delta^{(3)} ({\bf x}-{\bf x}_i(t)) , 
  \end{equation}
  \begin{equation}
    \varrho ({\bf x}, t; L) = \int {d{\bf y} \over L^3} \; 
    W \left( {|{\bf x} - {\bf y}| \over L} \right) 
    \varrho_{mic} ({\bf y}, t) .
  \end{equation}
\end{mathletters}
The physical interpretation of the field $\varrho ({\bf x}; L)$
follows straightforwardly from the properties of the smoothing window:
it is proportional to the number of particles contained within the
coarsening cell of size $\approx L$ centered at ${\bf x}$. A
microscopic peculiar-momentum density field and the corresponding
coarse-grained field can be defined in the same way:
\begin{mathletters}
  \label{momentum}
  \begin{equation}
    {\bf j}_{mic} ({\bf x}, t) = {m \over a(t)^3} 
    \sum_i {\bf u}_i(t) \; \delta^{(3)} ({\bf x}-{\bf x}_i(t)) ,
  \end{equation}
  \begin{equation}
    {\bf j} ({\bf x}, t; L) = \int {d{\bf y} \over L^3} \; 
    W \left( {|{\bf x} - {\bf y}| \over L} \right) 
    {\bf j}_{mic} ({\bf y}, t) .
\end{equation}
\end{mathletters}
One can introduce peculiar velocity fields ${\bf u}_{mic}$ and ${\bf
  u}$ by definition as ${\bf j} = \varrho \, {\bf u}$ and similarly
for ${\bf u}_{mic}$. The physical meaning of ${\bf u} ({\bf x}; L)$ is
also simple: it is the center-of-mass peculiar velocity of the
subsystem defined by the particles inside the coarsening cell of size
$\approx L$ centered at ${\bf x}$. Notice that ${\bf u}$ is {\em not}
obtained by coarse-graining ${\bf u}_{mic}$: from a dynamical point of
view, it is more natural to coarse-grain the momentum rather than the
velocity, since the former is an additive quantity for a system of
particles. Finally, one can define peculiar gravitational acceleration
fields ${\bf w}_{mic}$ and ${\bf w}$ through a coarse-graining of the
force:
\begin{mathletters}
  \label{acceleration}
  \begin{equation}
    \varrho_{mic} {\bf w}_{mic} ({\bf x}, t) = {m \over a(t)^3} 
    \sum_i {\bf w}_i(t) \; \delta^{(3)} ({\bf x}-{\bf x}_i(t)) ,
  \end{equation}
  \begin{equation}
    \varrho \, {\bf w} ({\bf x}, t; L) = \int {d{\bf y} \over L^3} \; 
    W \left( {|{\bf x} - {\bf y}| \over L} \right) 
    \varrho_{mic} {\bf w}_{mic} ({\bf y}, t) . 
  \end{equation}
\end{mathletters}
The field ${\bf w} ({\bf x})$ has the physical meaning of the
center-of-mass peculiar gravitational acceleration of the subsystem
defined by the coarsening cell at ${\bf x}$.

From these definitions and Eqs.\ (\ref{newton}a) and (\ref{newton}b),
it is straightforward to derive the evolution equations obeyed by the
coarse-grained fields $\varrho$ and ${\bf u}$ (from now on,
$\partial/\partial t$ is taken at constant ${\bf x}$, and $\nabla$
means partial derivative with respect to ${\bf x}$):
\begin{mathletters}
  \label{hydro}
  \begin{equation}
    \frac{\partial \varrho}{\partial t} + 3 H \varrho = - {1 \over a} 
    \nabla \cdot (\varrho \, {\bf u}) ,
  \end{equation}
  \begin{equation} 
    \frac{\partial (\varrho \, {\bf u})}{\partial t} + 4 H \varrho \, 
    {\bf u} = \varrho \, {\bf w} - 
    {1 \over a} \nabla \cdot (\varrho \, {\bf u} \, {\bf u} + \Pi) ,
  \end{equation}
\end{mathletters}
where a new second-rank tensor field has been defined:
\begin{equation}
  \label{veldis}
  \Pi ({\bf x}, t; L) = \int {d{\bf y} \over L^3} \; 
  W \left( {|{\bf x} - {\bf y}| \over L} \right)
  \varrho_{mic} ({\bf y}, t) \, 
  [{\bf u}_{mic}({\bf y}, t) - {\bf u}({\bf x}, t; L)] 
  [{\bf u}_{mic}({\bf y}, t) - {\bf u} ({\bf x}, t; L)] .
\end{equation}
The field $\Pi ({\bf x})$ is due to the velocity dispersion, i.e., to
the fact that the particles in the coarsening cell have in general a
velocity different from that of the center of mass. The trace of $\Pi$
is proportional to the internal kinetic energy of the coarsening cell,
that is, the total kinetic energy of the particles in the reference frame
of the center of mass.

The physical meaning of Eqs.\ (\ref{hydro}a) and (\ref{hydro}b) is
simple: they are just balance equations, stating mass conservation and
momentum conservation, respectively. The term $\nabla \cdot \Pi$
represents a {\em kinetic} pressure due to the exchange of particles
between neighboring coarsening cells (just like in the ideal gas) and
it has the same physical origin as the convective term $\nabla \cdot
(\varrho \, {\bf u} \, {\bf u})$, i.e., a nonlinear mode-mode coupling
of the velocity field. The difference is that the convective term
couples only modes on scales $> L$, while the velocity dispersion term
codifies the dynamical effect of the coupling of the modes on scales
$> L$ with the modes on scales $< L$. The term $\varrho \, {\bf w}$
codifies the gravitational interaction between the coarsening cells
and it is shown later that it can be split in a similar manner into a
contribution due to the large scales and another due to the coupling
of the large scales with the small ones. Although Eqs.\ (\ref{hydro})
look similar to the ordinary hydrodynamic equations, there is the
important difference that these equations are {\em exact}: as one
changes the smoothing length, the fields $\varrho$, ${\bf u}$, ${\bf
  w}$, $\Pi$ change but in such a way that the equations remain the
same (for example, upon increase of the smoothing length, part of the
dynamical effect described by $\nabla \cdot (\varrho \, {\bf u} \,
{\bf u})$ is shifted towards $\nabla \cdot \Pi$). This property is
reflected in that the equations are not an autonomous system for
$\varrho$ and ${\bf u}$. In fact, they are just the first ones of an
infinite hierarchy, as can be checked by computing the evolution
equations for the fields ${\bf w}$ and $\Pi$ (see Eq.\ (\ref{Pieq}),
for example). To obtain a useful set of equations, it is necesary to
truncate this hierarchy by looking for a functional dependence of
${\bf w}$ and $\Pi$ on $\varrho$ and ${\bf u}$. This will be the task
of the next section.

\section{The large-scale expansion}
\label{secsmallk}

In this section I discuss the closure of the hydrodynamic hierarchy at
the level of Eqs.\ (\ref{hydro}). When the particle interaction is
dominated by a fast-decaying (``collisional'') force, as in normal
fluids, this truncation is achieved by the assumption of local
equilibrium (see, e.g., Refs.\ \cite{IrKi50,Bale91}). In this case, the
interaction determines a privileged smoothing scale $L$ and it drives
the coarsening cells of size $L$ towards an approximate internal
thermal equilibrium, as if isolated from each other. The evolution of
the large scales ($\gg L$) is then ruled by the interaction between
neighboring coarsening cells. In this way, for example, Eq.\ 
(\ref{hydro}b) becomes Navier-Stokes' equation.

One cannot, however, apply this approach to the problem in hand: a
system with a long-ranged, unshielded interaction such as gravity
does not obey the usual thermodynamics (see, e.g., the brief review in
Ref.\ \cite{Hut97} and the more technical remarks in Ref.\ 
\cite{Bale91}). Moreover, this long range implies that the
interaction is self-similar and does not pick up itself a favored
coarsening length. One must therefore make use of a different approach
to close the hydrodynamic hierarchy.

For this purpose, I introduce the {\em large-scale expansion}, which
relies on the assumption that, in the context of cosmological
structure formation, the evolution of the large scales is weakly
influenced by what is going on in the small scales. This assumption is
further discussed in Sec.\ \ref{secconclusion}. Here I show how to
formulate this idea in order to write the fields ${\bf w}$ and $\Pi$
in Eqs.\ (\ref{hydro}) in terms of $\varrho$ and ${\bf u}$. Let a
tilde denote the Fourier transform of any field:
\begin{mathletters}
  \label{fourier}
  \begin{equation}
    \tilde{\phi} ({\bf k}) = \int_V d{\bf x} \; e^{i {\bf k} \cdot {\bf x}} 
    \, \phi ({\bf x}) ,
  \end{equation}
  \begin{equation}
    \phi({\bf x}) = {1 \over V} \sum_{\bf k} 
    e^{-i {\bf k} \cdot {\bf x}} \, \tilde{\phi} ({\bf k}) , 
  \end{equation}
\end{mathletters}  
where $V$ denotes the volume of periodicity for Eqs.\ (\ref{newton}).
Then the definition (\ref{density}b) can be written as
$\tilde{\varrho}({\bf k}; L) = \tilde{W}(L \, {\bf k}) \,
\tilde{\varrho}_{mic}({\bf k})$. Making use of the Taylor-expansion
(\ref{taylorwin}) and formally inverting it, one gets:
\begin{mathletters}
  \label{micdens}
  \begin{equation}
    \tilde{\varrho}_{mic}({\bf k}) = [ 1 + {1 \over 2} B \, (L \, k)^2 + 
    o(L \, k)^4 ] \, \tilde{\varrho}({\bf k}; L) , 
  \end{equation}
  \begin{equation}
    \varrho_{mic}({\bf x}) = [ 1 - {1 \over 2} B \, (L \, \nabla)^2 + 
    o(L \, \nabla)^4 ] \, \varrho({\bf x}; L) .
  \end{equation}
\end{mathletters}
Consider first the field ${\bf w}$. Fourier-transforming its
definition (\ref{acceleration}) and using Eqs.\ (\ref{newton}c) and
(\ref{newton}d), one finds:
\begin{equation}
  \label{wfourier}
  \widetilde{\varrho \, {\bf w}} \, ({\bf k}; L) = - {4 \pi G a \over V} 
  \sum_{{\bf q} \neq {\bf 0}} \, {i {\bf q} \over q^2} 
  \, \tilde{\varrho}_{mic}({\bf q}) \, \tilde{\varrho}_{mic}({\bf k}-{\bf q}) 
  \, \tilde{W}(L \, k) .
\end{equation}
This expression collects the mode-mode coupling via gravity of the
scales $\sim k^{-1}$ with every other scale. Take now a large scale
$k^{-1} \gg L$: the assumption of weak coupling to the small scales
then means that in the summation in Eq.\ (\ref{wfourier}), the main
contribution arises also from the large scales, $q^{-1}, |{\bf k}-{\bf
  q}|^{-1} \gg L$. One is then justified to insert the expansions
(\ref{micdens}a) and (\ref{taylorwin}) into Eq.\ (\ref{wfourier}).
Transforming back into real space yields the result:
\begin{mathletters}
  \label{wsmallk}
  \begin{equation}
    \varrho \, {\bf w} = \varrho \, {\bf w}^{mf} + 
    B L^2 (\nabla \varrho \cdot \nabla) {\bf w}^{mf} + o(L \, \nabla)^4 ,
  \end{equation}
  \begin{equation}
    \nabla \cdot {\bf w}^{mf} = - 4 \pi G a \, ( \varrho - \varrho_b ) , 
  \end{equation}
  \begin{equation}
    \nabla \times {\bf w}^{mf} = {\bf 0} .
  \end{equation}
\end{mathletters}
Thus, the unknown field ${\bf w}$ has been written as a functional of
the coarse-grained density field. The combination $L \, \nabla$
($\leftrightarrow L \, k$) can be {\it formally} viewed as a parameter
which measures the influence of the small scales over the large scales
and which has been assumed to be small: hence the name large-scale
expansion. The field ${\bf w}^{mf}$, which can be called the {\it
  macroscopic} gravitational field, represents the gravitational field
created by the monopole moment of the matter distribution in the
coarsening cells, i.e., as if they had no spatial extension ($L=0$).
It obeys the equations that one would have naively guessed and we see
that this is not the whole story: there exist a correction due to the
higher multipole moments that can then be properly called {\em tidal}
correction and it represents the coupling of the large to the small
scales induced by gravity. Hence, this decomposition into a
macroscopic field and a tidal correction is analogous to the
decomposition in Eq.\ (\ref{hydro}b) into convection, $\varrho \, {\bf
  u} \, {\bf u}$, and velocity dispersion, $\Pi$.

The same procedure can be now applied to the field $\Pi$. Now, ${\bf
  j}_{mic} = \varrho_{mic} {\bf u}_{mic}$ obeys an expansion like
(\ref{micdens}b), but ${\bf u}_{mic}$ does not, because it is not
defined by a straightforward coarse-graining. What one has in this
case is slightly more involved:
\begin{equation}
  \label{micvel}
  {\bf u}_{mic} = {{\bf j}_{mic} \over \varrho_{mic}} = 
  \frac{[1 - {1 \over 2} B \, (L \, \nabla)^2 + o(L \, \nabla)^4 ] 
    \, {\bf j}}{[1 - {1 \over 2} B \, (L \, \nabla)^2 + 
    o(L \, \nabla)^4 ] \, \varrho} = \left[ 1 - {1 \over 2} B \, (L \, \nabla)^2 
  - B L^2 {\nabla \varrho \over \varrho} \cdot \nabla + o(L \, \nabla)^4 
  \right] \, {\bf u} .
\end{equation}
Because of this minor complication, it is simpler in this case to work
directly with the representation in real space, Eq.\ (\ref{veldis}),
into which the expansion (\ref{micvel}) and the one corresponding to
${\bf j}_{mic}$ are now inserted. Because the smoothing window
effectively restricts $|{\bf x}-{\bf y}| \lesssim L$, the fields
within the integral evaluated at point ${\bf y}$ can be consistently
Taylor-expanded around point ${\bf x}$ to order $(L \, \nabla)^4$.
The final result reads:
\begin{equation}
  \label{Pismallk}
  \Pi = B L^2 \, \varrho (\partial_i {\bf u}) 
  (\partial_i {\bf u}) + o(L \, \nabla)^4 ,
\end{equation}
where a summation over the index $i$ is implied. Taking this result
and (\ref{wsmallk}) into the hydrodynamic hierarchy (\ref{hydro}), and
dropping the assumed small corrections $o(L \, \nabla)^4$, I finally
obtain
\begin{mathletters}
  \label{hydrosmallk}
  \begin{equation}
    \frac{\partial \varrho}{\partial t} + 3 H \varrho = - {1 \over a} 
    \nabla \cdot (\varrho \, {\bf u}) ,
  \end{equation}
  \begin{equation} 
    \frac{\partial (\varrho \, {\bf u})}{\partial t} + 4 H \varrho \, 
    {\bf u} = \varrho \, {\bf w}^{mf} - 
    {1 \over a} \nabla \cdot (\varrho \, {\bf u} \, {\bf u}) + 
    B L^2 \, \left\{ (\nabla \varrho \cdot \nabla) {\bf w}^{mf} - 
      {1 \over a} \nabla \cdot [\varrho (\partial_i {\bf u}) 
      (\partial_i {\bf u})] \right\} ,
  \end{equation}
  \begin{equation}
    \nabla \cdot {\bf w}^{mf} = - 4 \pi G a \, ( \varrho - \varrho_b ) , 
  \end{equation}
  \begin{equation}
    \nabla \times {\bf w}^{mf} = {\bf 0} .
  \end{equation}
\end{mathletters}
This is an autonomous system of equations for the two coarse-grained
fields $\varrho$ and ${\bf u}$. Compared to the hydrodynamic equations
of a normal fluid, we see that, because of the long range of the
interaction, its contribution $\varrho \, {\bf w}$ to the equation for
momentum conservation cannot be written as the divergence of a tensor,
i.e., as a nonideal correction to the kinetic contribution
represented by $\Pi$. Also the expression to lowest-order for $\Pi$
is another evident difference. In the next sections, I explore the
dynamical evolution described by Eqs.\ (\ref{hydrosmallk}).

\section{The dust model}
\label{secdust}

\begin{figure}
  \centerline{\includegraphics*[width=8cm]{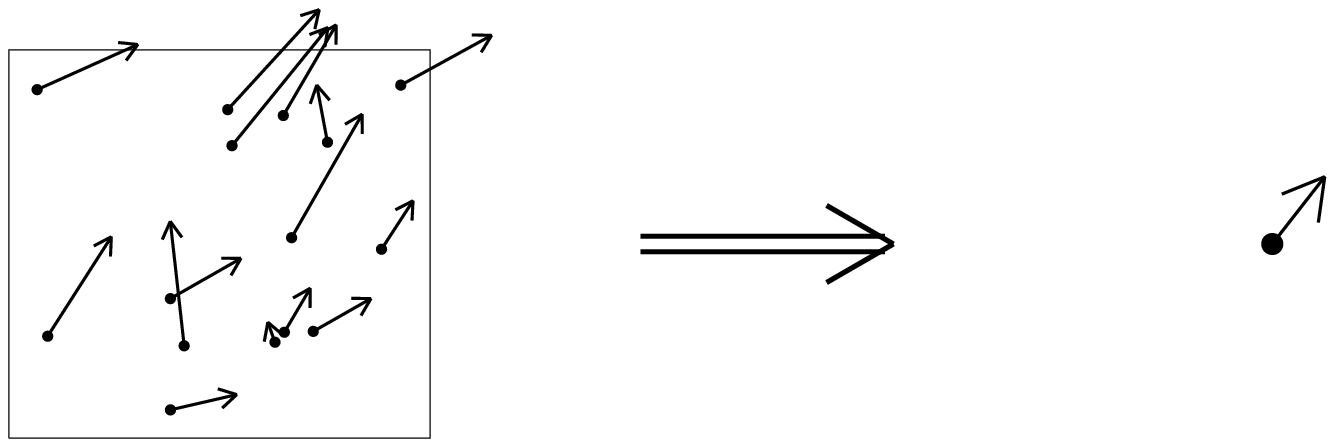}}
  \caption{
    The dust model amounts to neglecting altogether the internal
    structure (velocity dispersion and spatial extension) of the
    coarsening cell: it is approximated by a ``big particle'' located
    at the center of the cell, whose mass is that contained within it
    and whose velocity is the center-of-mass velocity.}
  \label{figdust}
\end{figure}

In this section I consider the lowest order in the large-scale
expansion. This corresponds to formally setting $L=0$ in Eq.\ 
(\ref{hydrosmallk}b), thus yielding:
\begin{mathletters}
  \label{dust}
  \begin{equation}
    \frac{\partial \varrho}{\partial t} + 3 H \varrho + {1 \over a} 
    \nabla \cdot (\varrho \, {\bf u}) = 0 ,
  \end{equation}
  \begin{equation} 
    \frac{\partial {\bf u}}{\partial t} + H {\bf u} + 
    {1 \over a} ({\bf u} \cdot \nabla) {\bf u} = {\bf w}^{mf} ,
  \end{equation}
  \begin{equation}
    \nabla \cdot {\bf w}^{mf} = - 4 \pi G a \, ( \varrho - \varrho_b ) , 
  \end{equation}
  \begin{equation}
    \nabla \times {\bf w}^{mf} = {\bf 0} .
  \end{equation}
\end{mathletters}
This is the popular and throughly studied {\it dust model} (see, e.g.,
Refs.\ \cite{Peeb80,Padm93,SaCo95}). The large-scale expansion
provides a clear picture of the approximations leading from Eqs.\ 
(\ref{newton}) to this model: by setting $L=0$ one assumes that the
coarsening cells can be thought of as ``big particles'' lacking
completely an internal structure (Fig.\ \ref{figdust}). This implies
neglecting (i) the velocity dispersion $\Pi$ compared to the
convection term $\varrho \, {\bf u} \, {\bf u}$; and (ii) the spatial
extension of the cells and thus the tidal correction compared to the
macroscopic gravitational field ${\bf w}^{mf}$.

It will be useful to briefly review some results for the dust model.
In the linear regime of small fluctuations about the homogeneous
cosmological background, Eqs.\ (\ref{dust}) can be linearized and the
resulting set of linear differential equations solved. In particular,
in the long-time limit (but still within the linear approximation),
the peculiar velocity and the gravitational acceleration are related
by the condition of parallelism:
\begin{equation}
  \label{parallelism}
  {\bf w}^{mf} ({\bf x}, t) = F(t) \, {\bf u} ({\bf x}, t) , 
\end{equation}
with the function
\begin{equation}
  F(t) = 4 \pi G \varrho_b(t) \, {b(t) \over \dot{b}(t)} > 0 ,
\end{equation}
where $b(t)$ is the growing mode of the (small) density contrast,
i.e., the growing solution of the equation $\ddot{b}+2 H \dot{b}-4 \pi
G \varrho_b b = 0$.

The linear evolution eventually breaks down due to the growth of
inhomogeneities by gravitational instability. The solution to the
fully nonlinear equations (\ref{dust}) is not known, but a successful
approximation in this regime is the {\em Zel'dovich approximation} (ZA
hereafter) \cite{SaCo95,Zeld70,ShZe89}, which turns out to be an exact
solution of Eqs.\ (\ref{dust}) for some highly symmetric
configurations. In the most general case, it can be understood as the
extrapolation of the parallelism condition (\ref{parallelism}) into
Eq.\ (\ref{dust}b) \cite{bib.parallel}. More properly, considering the
way Eqs.\ (\ref{dust}) were derived, the actual approximation is the
{\em truncated} ZA \cite{bib.tza}, since the density and velocity
fields have been smoothed on a scale $L$. Comparison with N-body
simulations \cite{bib.tza} shows that the truncated ZA performs
substantially better than the original ZA: this is no wonder within
the present coarse-graining formalism, since the particles that must
obey the ``dust'' evolution (and thus be moved according to the ZA)
are not the N-body particles of the simulations, which follow Eqs.\ 
(\ref{newton}), but rather the ``big particles'' which represent the
coarsening cells.

If one introduces a rescaled velocity field ${\bf v} = {\bf u} /a
\dot{b}$, and uses $b(t)$ as the new temporal variable, then Eq.\ 
(\ref{dust}b) becomes in the ZA:
\begin{mathletters}
  \label{zeldovich}
  \begin{equation}
    \frac{\partial {\bf v}}{\partial b} + ({\bf v} \cdot \nabla) 
    {\bf v} = 0 , 
  \end{equation}
  \begin{equation}
    \nabla \times {\bf v} = {\bf 0} ,  
  \end{equation}  
\end{mathletters}
together with the irrotationality constraint following from Eqs.\ 
(\ref{parallelism}) and (\ref{dust}d). Hence, the problem reduces to
the curl-free evolution of a fluid under no forcing at all. The
solution to this equation is then inserted into the continuity
equation (\ref{dust}a) to yield the evolved density field. As is
well-known, however, the ZA has the problem of giving rise to
singularities in the fields, mainly sheet-like singularities or
``pancakes'' \cite{ShZe89} at which $\nabla \cdot {\bf v} \rightarrow
- \infty$ and $\varrho \rightarrow + \infty$. Beyond this moment, the
ZA ceases to be valid.  The reason lies in the nonlinear, convective
term in Eq.\ (\ref{zeldovich}a), which deforms the initial velocity
field and generically leads to a multivalued velocity field ({\it
  shell crossing} in the cosmological literature). This feature is
likely not exclusive to the ZA but a generic property of the fully
nonlinear dust model, Eqs.\ (\ref{dust}), as checked, e.g., by the
application of Lagrangian perturbation techniques (see \cite{SaCo95}
and refs.  therein).  The generation of singularities by the
convective term is not prevented by the gravitational acceleration
${\bf w}^{mf}$, which, on the contrary, favors this process. The
emergence of singularities signals the unsuitability of the dust model
beyond that time: in the next section I investigate how the correction
to dust following from the large-scale expansion regularizes the
singularities.

\section{The adhesion model}
\label{secadhesion}

The singularities of the dust model arise because the approximation of
the coarsening cells as ``big particles'' is bad when the density of
these ``particles'' is large (formally infinite): the interaction
between them is no longer the simple macroscopic gravitational force
$\varrho \, {\bf w}^{mf}$, but it becomes more and more dependent on
the internal structure of the coarsening cells.  Therefore, a first
step to improve the dust model is to take into account the first-order
correction following from the large-scale expansion and study Eqs.\ 
(\ref{hydrosmallk}).

Obviously, the nonlinearities represented by this correction make it
even more difficult to solve these equations than in the case of the
dust model. Hence, in order to understand the effects of the
correction on the dust evolution, I introduce some further simplifying
assumptions. The idea is to consider the limit $B L^2 \rightarrow 0^+$
in Eqs.\ (\ref{hydrosmallk}), so that the correction is irrelevant
everywhere {\em except at those places where the dust evolution would
  predict a singularity}. Therefore, the evolution will follow the ZA
almost everywhere and the parallelism approximation
(\ref{parallelism}) will be good: this implies in particular, via
Eqs.\ (\ref{hydrosmallk}c) and (\ref{hydrosmallk}d), that ${\bf u}$ is
also curl-free and that $\nabla \cdot {\bf u} \propto -
(\varrho-\varrho_b)$. Since the correction to dust in Eq.\ 
(\ref{hydrosmallk}b) will be effective only near the singularities and
these correspond in the ZA predominantly to pancakes, the correction
can be computed under the approximation of a local plane-parallel
collapse: $\varrho$ and ${\bf u}$ change only along the direction
${\bf n}$ of the local plane-parallel collapse and $\partial_i u_j
\approx (\nabla \cdot {\bf u}) \, n_i n_j$. Making use of these
approximations, the correction term can be simplified as follows:
\begin{mathletters}
  \label{correction}
  \begin{equation}
    {1 \over \varrho} (\nabla \varrho \cdot \nabla) {\bf w}^{mf} 
    \approx {F \over \varrho} (\nabla \cdot {\bf u}) \nabla \varrho 
    \approx F \nabla^2 {\bf u} 
    \quad (\propto - \nabla \varrho) ,
  \end{equation}
  \begin{equation}
    {1 \over a \varrho} \nabla \cdot [\varrho (\partial_i {\bf u}) 
    (\partial_i {\bf u})] \approx
    {1 \over a \varrho} \nabla [\varrho (\nabla \cdot {\bf u})^2] 
    \approx {3 \over a} (\nabla \cdot {\bf u}) \nabla^2 {\bf u} 
    \quad (\propto - \varrho \nabla \varrho) ,
  \end{equation}
\end{mathletters}
where it has been taken into account that the correction will be
effective only at those places where $\varrho \sim (-\nabla \cdot {\bf
  u}) \rightarrow +\infty$. In this limit, the contribution
(\ref{correction}b) due to the velocity dispersion will be dominant
over that of the tidal correction (\ref{correction}a). Inserting this
result in Eq.\ (\ref{hydrosmallk}b) and in terms of the variables ${\bf
  v}$ and $b$ previously introduced, I finally arrive at the following
model:
\begin{mathletters}
  \label{adhesion}
  \begin{equation}
    \frac{\partial {\bf v}}{\partial b} + ({\bf v} \cdot \nabla) {\bf v} 
    = \nu \, |\nabla \cdot {\bf v}| \, \nabla^2 {\bf v} 
    \qquad (\nu = 3 B L^2 \rightarrow 0^+) , 
  \end{equation}  
  \begin{equation}
    \nabla \times {\bf v} = {\bf 0} ,
\end{equation}  
\end{mathletters}
This is a closed set of equations for the velocity field: once solved,
this field is used to compute the evolved density field from the
continuity equation (\ref{hydrosmallk}a), much in the same way as with
the ZA. The coefficient $\nu$ was called the {\it gravitational
  multistream} viscosity in Ref.\ \cite{BDP99}. Because it multiplies
the highest-order derivative in Eq.\ (\ref{adhesion}a), one cannot
simply drop this term in the limit $\nu \rightarrow 0^+$, but one must
rather apply the techniques of the boundary-layer theory to study this
limit. The mathematical handling is collected in the next subsection;
here I simply quote the conclusion that, as expected, the correction
regularizes the pancake-like singularities predicted by the dust
model, which become robust structures, where more and more mass gets
``adhered''. Another conclusion is that this ``adhesive'' behavior is
rather insensitive to the detailed functional dependence on $\nabla
\cdot {\bf v}$ of the factor in front of $\nabla^2 {\bf v}$: hence,
the same behavior arises in particular if one simply drops the
$|\nabla \cdot {\bf v}|$ to obtain a linear correction. But then, one
recovers the original adhesion model
\cite{SaCo95,ShZe89,bib.adhesion}, which was introduced as a
phenomelogical (and, when compared against N-body simulations
\cite{bib.adhesion}, very successful) correction to the ZA to go
beyond the epoch of formation of singularities.

\begin{figure}
  \centerline{\includegraphics*[width=8cm]{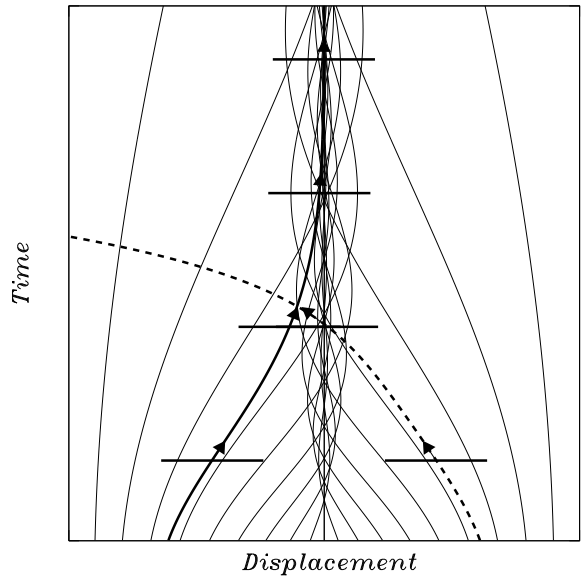}}
  \caption{
    Sketch of the particle trajectories forming a pancake, based on
    N-body simulations \protect\cite{Buch96}. The horizontal segment
    represents a coarsening cell.  The dashed trajectory corresponds
    to the dust model and its extrapolation (by the ZA) beyond the
    singularity; it is unable to reproduce the stabilization of the
    pancake. The solid trajectory represents the adhesion model and is
    more realistic.  There is no difference initially, but at the
    pancake the velocity dispersion becomes very large and produces an
    effective ``adhesive'' forcing that corrects the dust evolution.}
  \label{figadh}
\end{figure}

In the framework of the large-scale expansion, the physical
explanation of the adhesive behavior is clear (Fig.\ \ref{figadh}).
The corrections to dust (\ref{correction}) behave as a force in the
opposite direction to the density gradient, so that in the
neighborhood of a pancake there is a competition between the inflow of
matter driven by the gravitational attraction and this ``repulsion'',
which prevents the formation of a singularity. The dominant
contribution as $\varrho \rightarrow +\infty$ is the correction
(\ref{correction}b) due to the velocity dispersion and represents the
conversion of ``coherent'' streaming kinetic energy into
``disordered'' internal kinetic energy within the coarsening cells
\cite{BDP99,BuDo98,Buch98} (the mechanism has the same origin as the
pressure and viscous-like forces in a gas). The subdominant
contribution (\ref{correction}a) from the tidal correction also
exhibits this ``repulsive'' behavior, and means that the macroscopic
field ${\bf w}^{mf}$ overestimates the gravitational attraction
between coarsening cells.

\subsection{Application of the boundary-layer theory}
\label{secblayer}

In this subsection I apply the theory of boundary layers \cite{BeOr78}
to the model (\ref{adhesion}). The physical picture is that one can
study the evolution almost everywhere setting $\nu=0$. However, there
arise regions (``shocks'') spatially well separated from each other
and of vanishing volume (as $\nu \rightarrow 0^+$) where ${\bf v}$ has
a discontinuity, $|\nabla {\bf v}|$ diverges and the effect of the
term multiplied by $\nu$ must be taken into account. The fact that
these shocks correspond mainly to plane-parallel structures greatly
simplifies the technical handling, since it reduces the problem to the
solution of an ordinary differential equation. 

The purpose is to study Eqs.\ (\ref{adhesion}) in the vicinity of a
pancake, so that one takes a point at the pancake and introduces a new
coordinate system moving rigidly with it and with one of the axis
normal to the pancake. In this new noninertial reference frame Eq.\ 
(\ref{adhesion}a) must be appended with a term ${\bf A}$ which
collects the acceleration due to the inertial forces, and the velocity
${\bf v}$ is now understood as relative to this new reference frame.
New rescaled coordinates are defined as ${\bf x}'={\bf x}/\varepsilon$
and the idea is to take $\varepsilon \rightarrow 0^+$ in such a way
that the fields and their derivatives (and consequently also the
thickness of the pancake) remain finite in terms of the new
coordinates ${\bf x}'$ even in the limit $\nu \rightarrow 0^+$. (The
prime of the rescaled coordinates will be dropped hereafter to
simplify the notation). Let $n$, ${\bf x}_\parallel$ denote the
coordinates normal and parallel to the plane of the pancake,
respectively, and $v_n$, ${\bf v}_\parallel$ the corresponding
components of the velocity ${\bf v}$. Let $\nabla_\parallel$ denote
the gradient with respect to ${\bf x}_\parallel$. In terms of these
new coordinates, Eqs.\ (\ref{adhesion}) now read:
\begin{mathletters}
  \label{epsilonadhs}
  \begin{equation}
    \varepsilon \, \frac{\partial {\bf v}}{\partial b} - 
    \varepsilon \, {\bf A} - 
    \frac{d \varepsilon}{d b} \, \left( n \, \frac{\partial}{\partial n} + 
    {\bf x}_\parallel \cdot \nabla_\parallel \right) {\bf v} + 
    \left( v_n \, {\partial \over \partial n} + 
      {\bf v}_\parallel \cdot \nabla_\parallel \right) {\bf v} = 
    \varepsilon^{1-\gamma} \, \nu \left| \frac{\partial v_n}{\partial n} 
      + \nabla_\parallel \cdot {\bf v}_\parallel \right| ^{\gamma-2} 
    \, \left( {\partial^2 \over \partial n^2} + 
      \nabla_\parallel^2 \right) {\bf v} ,
  \end{equation}
  \begin{equation}
    \left( {\bf e}_n \, {\partial \over \partial n} + 
      \nabla_\parallel \right) \times ({\bf e}_n \, v_n + 
    {\bf v}_\parallel) = {\bf 0} ,
  \end{equation}
\end{mathletters}
where I have generalized the model (\ref{adhesion}) by allowing the
prefactor of $\nabla^2 {\bf v}$ to be an arbitrary function of $\nabla
\cdot {\bf v}$ whose asymptotic behavior as $\nabla \cdot {\bf v}
\rightarrow - \infty$ is characterized by the power $\gamma$ (the case
$\gamma=2$ recovers the original adhesion model). It has also been
considered the case that $\nu$ (and hence, the rescaling factor
$\varepsilon$) may depend on time. The motivation for this
generalization is twofold: (i) This kind of behavior corresponds to
the models discussed in Refs.\ \cite{BDP99,BuDo98,AdBu99}, where the
hydrodynamic hierarchy (\ref{hydro}) is closed by simply neglecting
tidal corrections altogether and assuming $\Pi_{ij} = \kappa
\varrho^\gamma \delta_{ij}$, $\kappa \rightarrow 0^+$. (ii) The
computation of the trace of $\Pi$ from N-body simulations also yields
this kind of behavior \cite{Domi00}.

The factor $\varepsilon$ must be chosen so as to absorb the explicit
dependence on $\nu$ in Eq.\ (\ref{epsilonadhs}a) and thus to render
the coefficient of the highest-order derivative of order unity, namely
$\varepsilon = \nu^{1/(\gamma-1)}$ and requiring $\gamma>1$, so that
$\varepsilon \rightarrow 0^+$. This latter constraint states that the
correction to dust must grow fast enough with $|\nabla {\bf v}|$ so as
to succesfully oppose the gravitational attraction and overcome the
formation of a singularity. Eqs.\ (\ref{epsilonadhs}) are now
simplified by keeping only the dominant terms in the limit
$\varepsilon \rightarrow 0^+$: first, one can set $\nabla_\parallel
\rightarrow {\bf 0}$, because in terms of the rescaled coordinates the
pancake looks like an infinite plane and the spatial variations along
it take place over an infinitely large lenght scale: only the
derivative normal to the pancake is relevant.  Also, the acceleration
terms $\partial {\bf v}/\partial b$ and ${\bf A}$ are finite in the
neighborhood of pancakes. Finally, $d \varepsilon / d b$ is of order
$\varepsilon$ at any finite time.  Therefore, Eqs.\ 
(\ref{epsilonadhs}) reduce to the following system of ordinary
differential equations:
\begin{mathletters}
  \label{blayer}
  \begin{equation}
    v_n \, {\partial v_n \over \partial n} = 
    \left| \frac{\partial v_n}{\partial n} \right| ^{\gamma-2} 
    \, {\partial^2 v_n \over \partial n^2} , 
  \end{equation}  
  \begin{equation}
    {\partial {\bf v}_\parallel \over \partial n} = {\bf 0} .
  \end{equation}  
\end{mathletters}
The physical interpretation of these equations is straightforward. The
tangential velocity, ${\bf v}_\parallel$, is smooth at the pancakes
and so constant in the rescaled coordinates. The normal velocity,
which looks discontinuous in the physical coordinates, is determined
in rescaled coordinates by the balance between the convective
transport towards the pancake (governed in turn by the gravitational
attraction of the pancake) and the outwards ``pressure'' due to the
velocity dispersion. 

\begin{figure}
  \centerline{\includegraphics*[width=15cm]{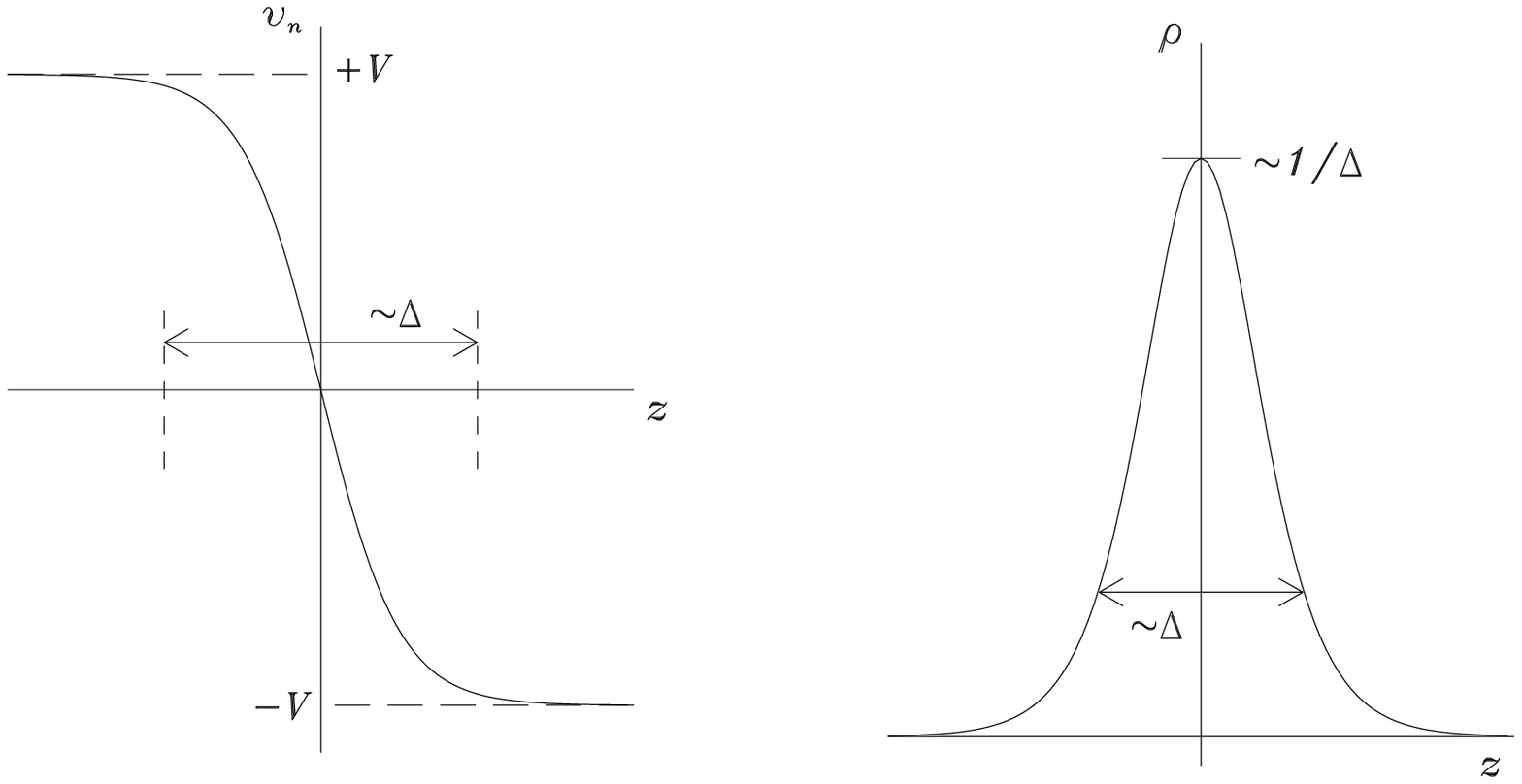}}
  \caption{
    Qualitative aspect of the coarse-grained density and velocity
    fields near a pancake according to Eqs.\ (\ref{generalsol}) and
    (\ref{denssolution}). $\Delta$ represents a measure of the pancake
    thickness.}
  \label{figpanc}
\end{figure}

The boundary conditions to be imposed to these equations are: (i)
$v_n(0) = 0$, meaning that the rescaled coordinate system is centered
at the pancake and moves with it; (ii) $\partial v_n/ \partial n \leq
0$ for any $n$, so that there is an inflow of matter towards the
pancake (and $\varrho \geq 0$, see Eq.\ (\ref{densitypancake})); (iii)
$\partial v_n / \partial n \rightarrow 0$ as $n \rightarrow \pm
\infty$, so that the derivative in physical coordinates is not
divergent outside of the pancake. Integrating twice Eq.\ 
(\ref{blayer}a) with these boundary conditions, one finds an implicit
solution $v_n (n)$:
\begin{equation}
  \label{generalsol}
  \int_0^{v_n(n) / V} d s \; (1 - s^2)^{1 \over 1-\gamma} = -
  {\varepsilon \, n \over \Delta} , \qquad 
  \Delta = \varepsilon \left( {\gamma-1 \over 2} \, 
    V^{3-\gamma} \right)^{1 \over 1-\gamma} ,
\end{equation} 
where $V>0$ is an integration constant. Knowning $v_n(n)$, one can
get an approximation to the pancake density profile by applying the
parallelism condition (\ref{parallelism}) together with Poisson
equation (\ref{hydrosmallk}c) in the limit $\varepsilon \rightarrow
0^+$:
\begin{equation}
  \label{densitypancake}
  \varrho = - {b \, \varrho_b \over \varepsilon} \, 
    \frac{\partial v_n}{\partial n} ,
\end{equation}
whose solution reads
\begin{equation}
  \label{denssolution}
  \varrho (n) = {b \, \varrho_b V \over \Delta} \left[ 1 - 
    {v_n (n)^2 \over V^2} \right] ^{1 \over \gamma-1} .
\end{equation}
A detailed study of the solution (\ref{generalsol}) leads to the
general picture shown in Fig.\ \ref{figpanc}: the fields remain
univalued at all times and no singularity arises. The parameter
$\Delta \rightarrow 0^+$ measures the pancake thickness in the
physical coordinates. This study also shows that for the cases
$\gamma>2$ the solution approaches so fast its asymptotic values at
large distances that in fact $\partial v_n/\partial n = 0$ at two
finite values of the normal coordinate, $n_+$ and $n_-=-n_+$. In such
cases, the solution (\ref{generalsol}) must be replaced beyond these
points by: $v_n (n)= (-) V$, $n<n_-$ ($>n_+$).

From the general solution (\ref{generalsol}), one can compute the
behavior of the solution $v_n (n = z/ \varepsilon)$ in the limit
$\varepsilon \rightarrow 0^+$ for a fixed value $z_0$ of the physical
coordinate $z = \varepsilon \, n$:
\begin{equation}
  \label{limit}
  v_n = \quad  V \text{  (if $z < z_0$),} \quad 0 \text{  (if $z = z_0$),} 
  \quad - V \text{  (if $z > z_0$),}
\end{equation}
independently of the value of $\gamma$ and the temporal dependence of
$\nu$. The conclusion is therefore that the adhesive-like behavior is
a robust property of the whole family of models (\ref{epsilonadhs})
parametrized by $\gamma$: whenever the dust evolution ($\nu=0$)
predicts a singularity, this must be replaced by the ``adhesive
prescription'' (\ref{limit}), which leads, via the continuity equation
(\ref{hydrosmallk}a), to a steady accretion of mass.

\section{Discussion}
\label{secconclusion}

I have derived and applied a hydrodynamic-like formulation for the
process of cosmological structure formation: this is a nontrivial
statement, because one could believe that the collisionless nature of
the basic model (\ref{newton}) renders a ``fluid'' description
impossible after shell-crossing, and that then one should employ a
different approach, e.g. the BBGKY hierarchy. I have shown, however,
that a ``fluid'' description is feasible after the breakdown of the
dust model: what makes a difference with ``down-to-Earth'' fluids is
how the closure of the hydrodynamic hierarchy (\ref{hydro}) is
achieved. For this pupose, I have introduced the large-scale
expansion, which builds on the assumption that the large-scale
evolution is insensitive to the small scales.  This assumption lies
behind many reasonings in the cosmological literature: for example,
behind the idea that on large scales the evolution follows a
Friedmann-Lema\^\i tre model, regardless of the small-scale
inhomogeneities, and also behind the confidence on cosmological N-body
simulations, where each N-body particle is so massive that it must
correspond in the real world to a full structure in its own, composed
of many smaller particles. The plausibility of the assumption can be
argued on the basis of the long range of gravity: the evolution can be
expected to be dominated by the large scales provided there is enough
large-scale power initially. (One can then also expect that the
validity of the large-scale expansion should depend on the initial
and boundary conditions). The good agreement with N-body simulations
of the truncated ZA and the adhesion model
\cite{bib.tza,bib.adhesion}, in which small-scale structure is
completely disregarded, can be viewed as a support of this
large-scale dominance.

From the derived hydrodynamic equations, I have shown how the dust
model arises as the lowest-order term in the large-scale expansion,
while the first-order correction gives rise to a model which can be
reduced to the adhesion model by further approximations. This
derivation provides a clear physical interpretation of the two models:
the ``particles'' which are assumed to follow the dust model are not
such but have an internal structure, and the ``interaction'' between
these ``particles'' due to its internal structure explains the
adhesive behavior. By applying boundary-layer techniques, I found in
particular that this behavior is a robust property of the model, in
the sense that it arises quite independently of the detailed
dependence of the velocity dispersion and the tidal correction on the
density and velocity fields.

It is interesting to compare this work with previous, related works
dealing also with a hydrodynamic-like formulation of large-scale
structure formation. In Refs.\ \cite{BDP99,DHMP99}, the case is
studied in which the hydrodynamic hierarchy is truncated by a {\em
  phenomenological} ansatz which writes the corrections to the
adhesion model as a stochastic term (a noise). The adhesion model, in
turn, was justified in Refs.\ \cite{BDP99,BuDo98,AdBu99} (see Ref.\ 
\cite{MTM99} for a relativistic generalization) by disregarding the
tidal correction altogether and by letting the velocity dispersion be
given as $\Pi_{ij} = \kappa \varrho^\gamma \delta_{ij}$. The
robustness of the adhesive property already mentioned explains why
this behavior also arises in the models that follow from this
truncation. It is particularly instructive to compare with the work in
Ref.\ \cite{BuDo98}, where a truncation relying on something else than
phenomenology was studied. I have been able to close the hydrodynamic
hierarchy with less restrictive assumptions and thus shown that the
adhesive behavior can still be recovered when assumptions (A2)
(isotropic $\Pi$) and (A4) and (A5) (further restrictions on the form
of $\Pi$) in Sec.\ 4 of Ref.\ \cite{BuDo98} are dropped. Assumption
(A1) seems in practice equivalent to the large-scale expansion, while
I also employed assumption (A3) (parallelism) in order to derive
adhesion-like models. But there is also a major improvement compared
to that work: I do not assume the mean-field approximation from the
outset (and thus the Vlasov equation, which was the starting point in
Ref.\ \cite{BuDo98}). In fact, I could derive an expression for the
tidal correction and show that it also behaves ``adhesively''. As
explained in more detail in App.\ \ref{apPi}, this correction is also
the origin for the discrepancy between expression (\ref{Pismallk}) and
the relationship $\Pi_{ij} = \kappa \varrho^{5/3} \delta_{ij}$ derived
in Ref.\ \cite{BuDo98}.

Finally, I have shown that the derivation of adhesion-like models
requires in principle further assumptions than the simple large-scale
expansion. Thus, one could improve on these models by relaxing those
assumptions and rather studying Eqs.\ (\ref{hydrosmallk}). An important
difference between the adhesion-like models and Eqs.\ 
(\ref{hydrosmallk}) is that the corrections to dust in the latter
generate vorticity, even if it is initially absent: this may be a
relevant feature when modelling galaxy formation. Another difference
is that now there is no need in principle to retain the condition $B
L^2 \rightarrow 0^+$: pancakes and other singularities are no longer
of vanishing volume, but have an inner structure whose evolution could
be studied with Eqs.\ (\ref{hydrosmallk}). For such purpose, the role
of the length $L$ must be better understood.

\acknowledgments
  
I would like to thank C. Beisbart, T. Buchert and H. Wagner for their
useful comments on the manuscript.

\appendix

\section{Smoothing windows}
\label{apwindow}

A {\it smoothing window} $W(z)$\cite{fn3} should behave as a {\it
  window} that defines a bounded region of space and as a {\it
  smoothing} filter that erases the structural details inside this
region. These conditions are implemented by requiring $W(z)$ and its
Fourier transform $\tilde{W}(q)$ to decay faster than any power as $z
\rightarrow \infty$ or $q \rightarrow \infty$, respectively. The
smoothing window acts as an integral kernel: if $\phi({\bf x})$ is a
given field, then the associated coarse-grained field $\phi_{L}({\bf
  x})$ over the scale $L$ and its Fourier transform $\tilde{\phi}_L
({\bf k})$ are given by
\begin{mathletters}
  \label{coarsegraining}
  \begin{equation}
    \phi_{L}({\bf x}) = {1 \over L^3} \int d{\bf y} \; W \left( 
      {|{\bf x} - {\bf y}| \over L} \right) \phi({\bf y}) , 
    \end{equation}
    \begin{equation}
      \tilde{\phi}_L ({\bf k}) = \tilde{W} (L \, {\bf k}) \, 
      \tilde{\phi}({\bf k}) .
    \end{equation}
\end{mathletters}
Therefore, the coarse-grained field at point ${\bf x}$ is just the
(weigthed) addition of the original field over a region of size
$\approx L$ around that point. Two more conditions are also required:
(i) $W(z=0)=1$, so that the contribution from the neighborhood of the
center of the window is unweighted; (ii) $\tilde{W}(q=0)=1$, so that
the coarse-grained field has the same large-scale structure as the
original field. This latter condition implies the following
normalization:
\begin{equation}
  \label{normalW}
  \int d{\bf x} \; W(|{\bf x}|) = 1 ,
\end{equation}
so that $\lim_{L \rightarrow 0} L^{3} W(|{\bf x}|/L) = \delta({\bf
  x})$, and $\lim_{L \rightarrow 0} \phi_{L}({\bf x}) = \phi({\bf
  x})$.

The condition on the decay of $\tilde{W}(q)$ for large $q$ prevents
the window $W(z)$ from having sharp borders, which thus becomes a
``fuzzy'' window. Otherwise, the coarse-grained field $\phi_{L}({\bf
  x})$ could change in a discontinuous manner as the center of the
window, ${\bf x}$, sweeps the system. In the same way, the fast decay
of $W(z)$ for large $z$ (which implies that $\tilde{W}(q)$ lacks sharp
borders in Fourier space) implies the property of spatial locality for
the coarse-grained field. This also guarantees the existence of the
Taylor-expansion of $\tilde{W}(q)$ at ${\bf q}={\bf 0}$ in the form:
\begin{equation}
  \label{taylorwin}
  \tilde{W}(q) = 1 - {1 \over 2} B \, q^2 + o(q^4) , 
\end{equation}
where the constant
\begin{equation}
  B = {1 \over 3} \int d{\bf z} \; z^2 \, W(z) = {4 \pi \over 3} 
  \int_0^{+\infty} d z \; z^4 \, W(z)
\end{equation}
is related to the quadrupole moment of the smoothing window.

The required constraints on the smoothing window exclude a step
function (either in real or in Fourier space). A useful function which
satisfies the constraints is a Gaussian: $W(z) = \exp (- \pi z^2)
\Rightarrow \tilde{W}(q) = \exp (- q^2 / 4 \pi)$.

\section{The evolution equation for $\Pi$}
\label{apPi}

In this appendix I study the evolution equation for the velocity
dispersion and show that (\ref{Pismallk}) is a solution of the
equation. The purpose is to compare with the result in Ref.\ 
\cite{BuDo98} for $\Pi$, also obtained from the evolution equation for
this tensor.

Starting from the definition (\ref{veldis}) and Eqs.\ (\ref{newton}a)
and (\ref{newton}b), one can obtain the following equation for the
temporal evolution of the velocity dispersion:
\begin{equation}
  \label{Pieq}
  \frac{\partial \Pi_{ij}}{\partial t} + 5 H \, \Pi_{ij} = 
  - {1 \over a} \partial_k (u_k \Pi_{ij}) 
  - {1 \over a} \Pi_{ik} \partial_k u_j
  - {1 \over a} \Pi_{jk} \partial_k u_i
  - {1 \over a} \partial_k {\cal L}_{ijk} + {\cal P}_{ij} ,
\end{equation}
where a summation over repeated indices is implied, and a second-rank
and a third-rank tensor fields have been defined:
\begin{mathletters}
  \label{PandL}
  \begin{eqnarray}
    {\cal P} ({\bf x}, t; L) = \int {d{\bf y} \over L^3} \; 
    W \left( {|{\bf x} - {\bf y}| \over L} \right)
    \varrho_{mic} ({\bf y}, t) \Big\{
      [{\bf u}_{mic}({\bf y}, t) - {\bf u}({\bf x}, t; L)] 
      [{\bf w}_{mic}({\bf y}, t) - {\bf w} ({\bf x}, t; L)] + 
      \nonumber \\
      \qquad\qquad\qquad\qquad\qquad\qquad\qquad\qquad\qquad
      \qquad\qquad
      + [{\bf w}_{mic}({\bf y}, t) - {\bf w}({\bf x}, t; L)] 
      [{\bf u}_{mic}({\bf y}, t) - {\bf u} ({\bf x}, t; L)]
    \Big\} ,
  \end{eqnarray}
  \begin{equation}
    {\cal L} ({\bf x}, t; L) = \int {d{\bf y} \over L^3} \; 
    W \left( {|{\bf x} - {\bf y}| \over L} \right)
    \varrho_{mic} ({\bf y}, t)
    [{\bf u}_{mic}({\bf y}, t) - {\bf u}({\bf x}, t; L)] 
    [{\bf u}_{mic}({\bf y}, t) - {\bf u} ({\bf x}, t; L)] 
    [{\bf u}_{mic}({\bf y}, t) - {\bf u}({\bf x}, t; L)] .
  \end{equation}
\end{mathletters}
The term ${\cal L}$ accounts for the change of velocity dispersion due
to the exchange of particles between the coarsening cells, while the
term ${\cal P}$ represents the change due to the gravitational
interaction. To get a physical picture, just consider the equation for
the trace of $\Pi$ (the internal kinetic energy): then ${\cal L}$
reduces to the equivalent of the kinetic contribution to the heat flux
in the usual hydrodynamics, while ${\cal P}$ becomes the power
performed by the gravitational interaction.

Compared to Eq.\ (4c) of \cite{BuDo98}, Eq.\ (\ref{Pieq}) contains the
extra term ${\cal P}$, and the reason is that this term drops if tidal
corrections are neglected. Indeed, if one performs a large-scale
expansion of ${\cal P}$ and ${\cal L}$ in the same way as explained in
Sec.\ \ref{secsmallk}, one obtains
\begin{mathletters}
  \begin{equation}
    {\cal P} = B L^2 \, \varrho [(\partial_i {\bf u}) 
    (\partial_i {\bf w}^{mf}) + (\partial_i {\bf w}^{mf}) 
    (\partial_i {\bf u})] + o(L \, \nabla)^4 , 
  \end{equation}
  \begin{equation}
    {\cal L} = o(L \, \nabla)^4 .
  \end{equation}
\end{mathletters}
When these expressions are inserted into Eq.\ (\ref{Pieq}), one can
check that the velocity dispersion tensor $\Pi$ given by
(\ref{Pismallk}) is indeed a solution.

In Ref.\ \cite{BuDo98}, Eq.\ (\ref{Pieq}) (with ${\cal P}=0$, as
explained) was solved for the trace of $\Pi$ by imposing the
constraint of shear-free flow: it is then found that $Tr \, \Pi =
\kappa \varrho^{5/3}$, where $\kappa$ is determined by the initial
velocity dispersion. This must be viewed as a solution valid only for
early times; when the tidal corrections grow, the term ${\cal P}$
becomes relevant: $\Pi$ eventually forgets its initial condition, due
to the term $5 H \, \Pi$ in Eq.\ (\ref{Pieq}), and becomes ``slaved''
to the density and velocity fields, as given by Eq.\ (\ref{Pismallk}).
This is a mechanism similar to the ``slaving'' represented by the
parallelism condition (\ref{parallelism}) in the linearized dust
evolution.

\newpage

\begin{references}

\bibitem{Peeb80}
  P.~J.~E. Peebles, {\em The Large-Scale Structure of the Universe} (Princeton
  University Press, Princeton, 1980).

\bibitem{Padm93}
  T. Padmanabhan, {\em Structure Formation in the Universe} (Cambridge University
  Press, Cambridge, 1993).
  
\bibitem{BuEh97}
  T. Buchert and J. Ehlers, Astron. Astrophys. {\bf 320},  1  (1997).

\bibitem{IrKi50}
  J.~H. Irving and J.~G. Kirkwood, J. Chem. Phys. {\bf 18},  817  (1950).

\bibitem{Bale91}
  R. Balescu, {\em Equilibrium and Nonequilibrium Statistical Mechanics} (John
  Wiley \& Sons, New York, 1991).
  
\bibitem{Hut97}
  P. Hut, Complexity {\bf 3},  38  (1997).
  
\bibitem{SaCo95}
  V. Sahni and P. Coles, Phys. Rep. {\bf 262},  1  (1995).
  
\bibitem{Zeld70}
  Y.~B. Zel'dovich, Astron. Astrophys. {\bf 5},  84  (1970).
  
\bibitem{ShZe89}
  S.~F. Shandarin and Y.~B. Zel'dovich, Rev. Mod. Phys. {\bf 61},  185  (1989).
  
\bibitem{bib.parallel} 
  S. Bildhauer and T. Buchert, Prog. Theor. Phys.  {\bf 86}, 653
  (1991); T. Buchert, Mon. Not. R. Astron. Soc. {\bf 254}, 729 (1992);
  L.~A. Kofman, in {\em IUAP Proceedings on Nucleosynthesis in the
    Universe}, edited by K. Sato (Kluwer:Dordrecht, 1991).
  
\bibitem{bib.tza}    
  P. Coles, A.~L. Melott, and S.~F. Shandarin, Mon. Not. R. Astron.
  Soc. {\bf 260}, 765 (1993); A.~L. Melott, T. Pellman, and S.~F.
  Shandarin, Mon. Not.  R. Astron. Soc. {\bf 269}, 626 (1994).
    
\bibitem{BDP99}  
  T. Buchert, A. Dom{\'\i}nguez, and J. P\'erez-Mercader, Astron.
  Astrophys. {\bf 349}, 343 (1999).
  
\bibitem{bib.adhesion}
  S.~N. Gurbatov, A.~I. Saichev, and S.~F. Shandarin, Mon. Not. R.
  Astron. Soc.  {\bf 236}, 385 (1989); D.~H. Weinberg, and J.~E. Gunn,
  Mon. Not. R.  Astron. Soc.  {\bf 247}, 260 (1990); L. Kofman, D.
  Pogosyan, S.~F.  Shandarin, and A.~L. Melott, Astrophys. J. {\bf
    393}, 437 (1992); A.~L. Melott, S.~F. Shandarin, and D.~H.
  Weinberg, Astrophys. J.  {\bf 428}, 28 (1994); B.~S. Sathyaprakash
  {\it et~al.}, Mon. Not. R.  Astron. Soc. {\bf 275}, 463 (1995).
  
\bibitem{BuDo98}
  T. Buchert and A. Dom{\'\i}nguez, Astron. Astrophys. {\bf 335},  395  (1998).

\bibitem{Buch98} 
  T. Buchert, in {\em From Stars to the Universe} (Annals of Shanghai
  Observatory, Shangai, 1998).

\bibitem{BeOr78}
  C.~M. Bender and S.~A. Orszag, {\em Advanced Mathematical Methods for
    Scientists and Engineers} (McGraw-Hill, New York, 1978).
  
\bibitem{AdBu99}
  S. Adler and T. Buchert, Astron. Astrophys. {\bf 343},  317  (1999).

\bibitem{Domi00}
  A. Dom{\'\i}nguez, submitted to Astron. Astrophys.
  
\bibitem{DHMP99} 
  A. Dom{\'\i}nguez {\it et~al.}, Astron. Astrophys. {\bf 344}, 27
  (1999).

\bibitem{MTM99}
  R. Maartens, J. Triginer and D. Matravers, Phys. Rev. D {\bf 60},
  103503 (1999).

\bibitem{fn3}
  Restricting the dependence to $z\equiv |{\bf z}|$ means a
  spherically symmetric window, so that the process of
  coarse-graining does not introduce any favored direction.
  
\bibitem{Buch96} T. Buchert, in {\em Proc. ``International School of
    Physics Enrico Fermi''}, edited by S. Bonometto, J. Primack, A.
  Provenzale (IOP Press: Amsterdam, 1996).
  
\end{references}
\end{document}